\documentclass{article}

\usepackage{icrctc07}

\title{Search for Dark Matter Annihilation in Draco with STACEE }
\shorttitle{Recent Observations of Draco}
\authors{D.D.~Driscoll$^a$,
J.~Ball$^b$, J.E.~Carson$^{b,1}$, C.E.~Covault$^a$, 
P.~Fortin$^c$, D.M.~Gingrich$^{d,e}$,
D.S.~Hanna$^f$, A.~Jarvis$^b$, J.~Kildea$^{f,2}$, T.~Lindner$^{f,3}$, 
C.~Mueller$^f$, R.~Mukherjee$^c$,
R.A.~Ong$^b$, K.~Ragan$^f$, D.A.~Williams$^g$, J.~Zweerink$^b$ }
\shortauthors{Driscoll, et al.}
\afiliations{
		$^{a}$  Dept.~of Physics, Case Western Reserve University,            Cleveland, OH 44106\\
		$^{b}$  Dept.~of Physics and Astronomy, Univ.~of 
                        California, Los Angeles, CA 90095\\
		$^{c}$  Dept.~of Physics and Astronomy, Barnard College, 
		     Columbia University, New York, NY 10027\\
		$^{d}$  Dept.~of Physics, Univ.~of Alberta, 
		     Edmonton, AB T6G 2G7\\
		$^{e}$  TRIUMF, Vancouver, BC V6T 2A3 Canada\\
		$^{f}$  Dept.~of Physics, McGill University, 
		     Montreal, QC H3A 2T8\\
		$^{g}$  Santa Cruz Inst. for Particle Physics,
		     Univ.~of California, Santa Cruz, CA 
                     95064\\
		$^{1}$  Current Address:  Stanford Linear Accelerator Center, 
		     Menlo Park, CA 94025\\
		$^{2}$  Current Address:  F.L. Whipple Obs., 
		     Harvard-Smithsonian Center for Astrophysics, 
                     Amado, AZ 85645\\
		$^{3}$  Current Address:  Dept.~of Physics and Astr.,
		Univ.~of British Columbia, Vancouver, BC 
}
\email{ddd3@po.cwru.edu}

\abstract{
For some time, the Draco dwarf spheroidal galaxy has garnered interest as a
possible source for the indirect detection of dark matter.  Its large
mass-to-light ratio and relative proximity to the Earth provide favorable
conditions for the production of detectable gamma rays from dark matter
self-annihilation in its core.  
The Solar Tower Atmospheric Cherenkov Effect Experiment (STACEE) is an
air-shower Cherenkov telescope located in Albuquerque, NM capable of
detecting gamma rays at energies above 100 GeV.  We present
the results of the STACEE observations of Draco during the 2005-2006
observing season totaling 10 hours of livetime after cuts.
}

\begin{document}
\maketitle

\section{Introduction}

Dark matter is thought to be an important component of the Universe
and research into its nature is actively pursued using a variety of
techniques.  Dark matter may be weakly interacting massive particles
(WIMPs) which would tend to accumulate at the bottom of gravitational
potential wells, such as galaxies, where they could undergo
self-annihilation processes. Depending on WIMP mass and branching
ratios, a measurable flux of high energy gamma rays could result.

The Draco dwarf spheroidal galaxy has long garnered interest as a
potential source of concentrated dark matter~\cite{Tyler}. Draco has
one of the highest known mass-to-light ratios ($M/L$), perhaps as high
as $500 M_\odot/L_\odot$~\cite{Wu}.  Current observations
are consistent with a cuspy density profile~\cite{Lokas},
which would enhance the gamma-ray production rate. Furthermore, since
Draco is a satellite of the Milky Way, its relative proximity to the
Earth ($d \sim 75~kpc$)\cite{Bonanos} might allow for the detection of
such gamma rays.

\section{STACEE Observations of Draco}

\begin{table*}[ht]
\begin{center}
\begin{tabular}{|l|r|r|r|r|} \hline
~						& ON events				& OFF events			& Excess 	& Significance 	\\ \hline
After Time Cuts	& 177498 		& 177273			& 225			& $+0.39\sigma$	\\
+ grid ratio Cut	& 3094 		& 3120 		& -26			& $-0.33\sigma$	\\ \hline
\end{tabular}
\caption{Data summary of STACEE observations of Draco during the 2005-2006 
	observing season, representing approximately $3.67\times10^{4}~s$ 
	of livetime.
\label{datasum}}
\end{center}
\end{table*}

\begin{figure}[bt]
	\begin{center}
	\includegraphics[width=0.45\textwidth]{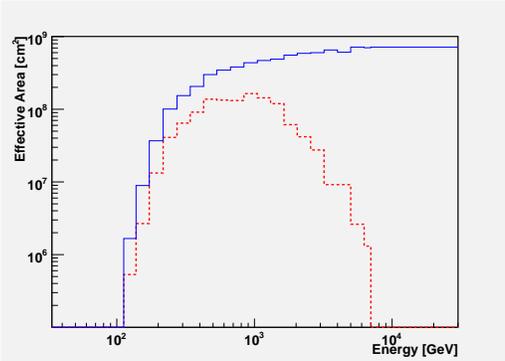}
	\caption{Effective area curves for STACEE observations of Draco.
		The blue (solid) line represents the STACEE effective area without cuts, the red (dashed)
		line represents the STACEE effective area including the grid-ratio cut.
	\label{effarea}}
	\end{center}
\end{figure}

The Solar Tower Atmospheric Cherenkov Effect Experiment (STACEE) is a
gamma-ray telescope operating at the National Solar Thermal Test
Facility (NSTTF) in Albuquerque, NM.  STACEE is a wavefront-sampling
Cherenkov telescope which uses 64 of the mirrors in the NSTTF
heliostat array for a total of $\sim 2400~m^2$ of collecting surface.
Cherenkov light from gamma-induced air showers is reflected off the
heliostats onto secondary mirrors on a tower on the south side of the
field.  These secondaries focus the light onto photomultiplier tubes
(PMTs) in such a way that each PMT sees the light from a single
heliostat.  Pulses from the PMTs are split, with one copy
discriminated and used in the formation of a trigger and the other
digitized using a 1 GS/s FADC.  The trigger selects showers that
deposit light evenly over the heliostat field, which favors those
showers initiated by gamma rays over those resulting from charged
cosmic rays, the most important background for the STACEE experiment.
For a more complete description of the STACEE experiment, see
\cite{Gingrich}.

The basic unit of observation for STACEE is the ``ON-OFF'' pair; 28 minutes
on-source and 28 minutes off-source.  Both observations view the same path
across the sky in local coordinates (altitude and azimuth), but separated by
30 minutes in celestial coordinates (right ascension).  The
off-source observation allows for a measurement of the local
background conditions.  We measure the significance of a measurement as
\begin{equation}
	\sigma = \frac{ON-OFF}{\sqrt{ON+OFF}} 
	\approx \frac{S}{\sqrt{2B}} \label{sig}
\end{equation} 
where $S$ is the signal and $B$ is the background.

STACEE observations of Draco total 35 ``ON-OFF'' pairs, of which approximately 10 hours
of livetime remain after excluding periods with bad weather and known
technical difficulties.  Our data set is summarized in Table \ref{datasum}.

\section{Data Analysis}

Our raw background trigger rate from cosmic rays is approximately 5
Hz.  In order to reduce this, we perform a grid-ratio cut which
preferentially removes hadron-induced showers.  This technique has
been used successfully by the CELESTE experiment~\cite{Brion} and our
implementation is described in more detail in~\cite{Kildea}.  A
simplistic description of the technique is that the ``smoothness'' of
a shower is measured by the height-to-width ratio ($H/W$) of the sum
of pulses from all 64 channels in the detector.  This quantity depends
on the relative timing of each FADC trace, which depends on the
assumed impact point of the shower core (i.e., the extrapolated shower
axis).  The grid-ratio cut is based on how sharply peaked the $H/W$
distribution as a function of assumed core position is.  Gamma-ray
showers, smooth and symmetric, are expected to produce narrower $H/W$
distributions than hadronic showers, which result in broader, clumpier
deposits of Cherenkov light.  Applied to our 2002-2004 Crab data, the
grid-ratio cut improves the detection significance from $4.8\sigma$ to
$8.1\sigma$\cite{Lindner}.

As seen in Table \ref{datasum}, we do not detect an excess gamma-ray signal
from Draco in our data set.  We derive an upper limit for the
flux from Draco given a measure of our detector response to a candidate
source spectrum.  We discuss two possible source spectra, an $E^{-2.2}$
power law (suggested by the gamma-ray flux from the galactic
center\cite{Hooper}) and a candidate dark matter spectrum taken from
Tyler\cite{Tyler}, with an energy-dependent shape given by
\begin{equation}
	\phi(x<1) = \frac{e^{-8x}}{x(x^{1.5} + 0.00014)}, 
	\label{tspec}
\end{equation}
where $ x = E / m_\chi c^2 $. This gives a sharp cutoff at the energy
corresponding to the candidate WIMP mass.

\begin{figure*}[tb]
	\begin{center}
	\includegraphics[width=0.45\textwidth]{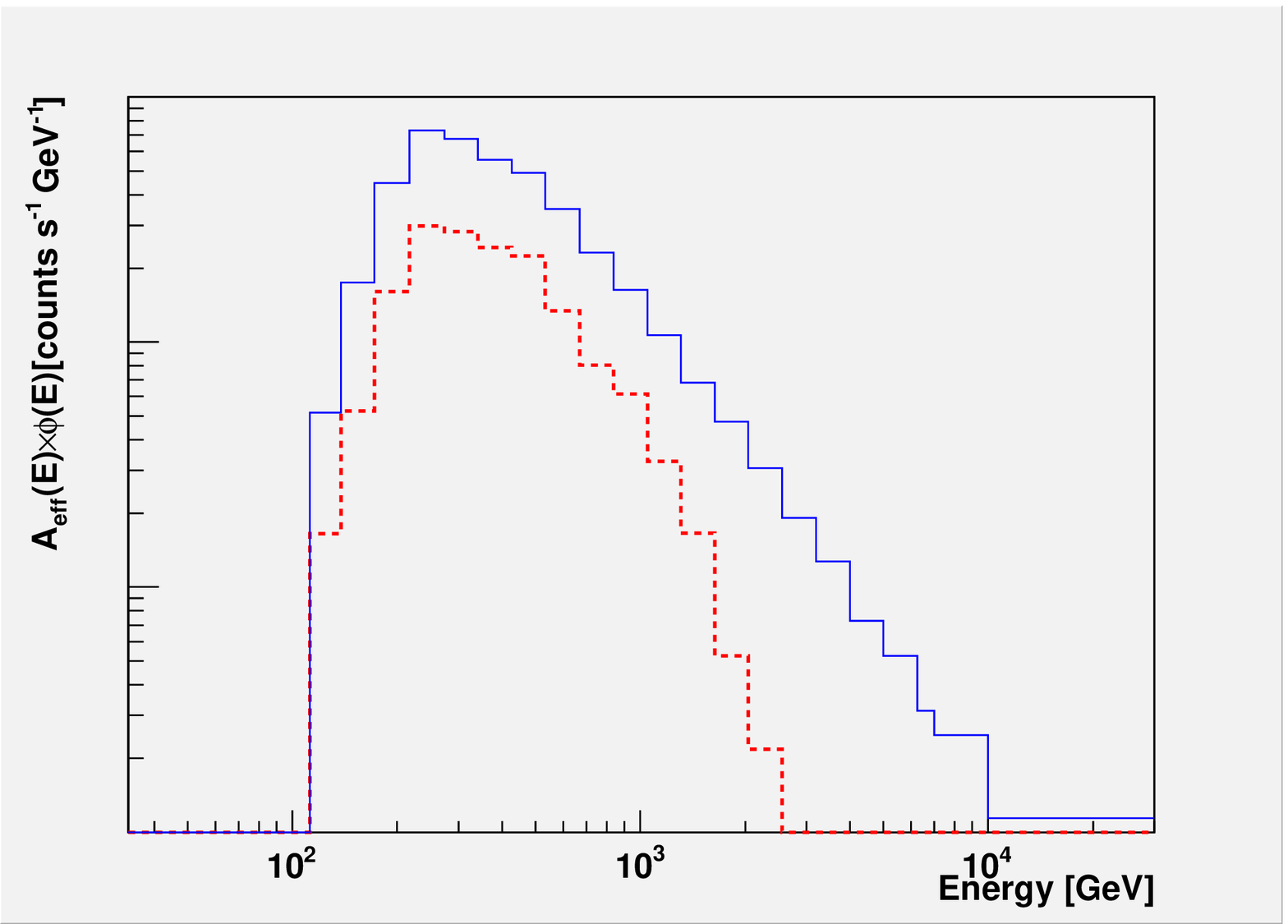}
	\includegraphics[width=0.45\textwidth]{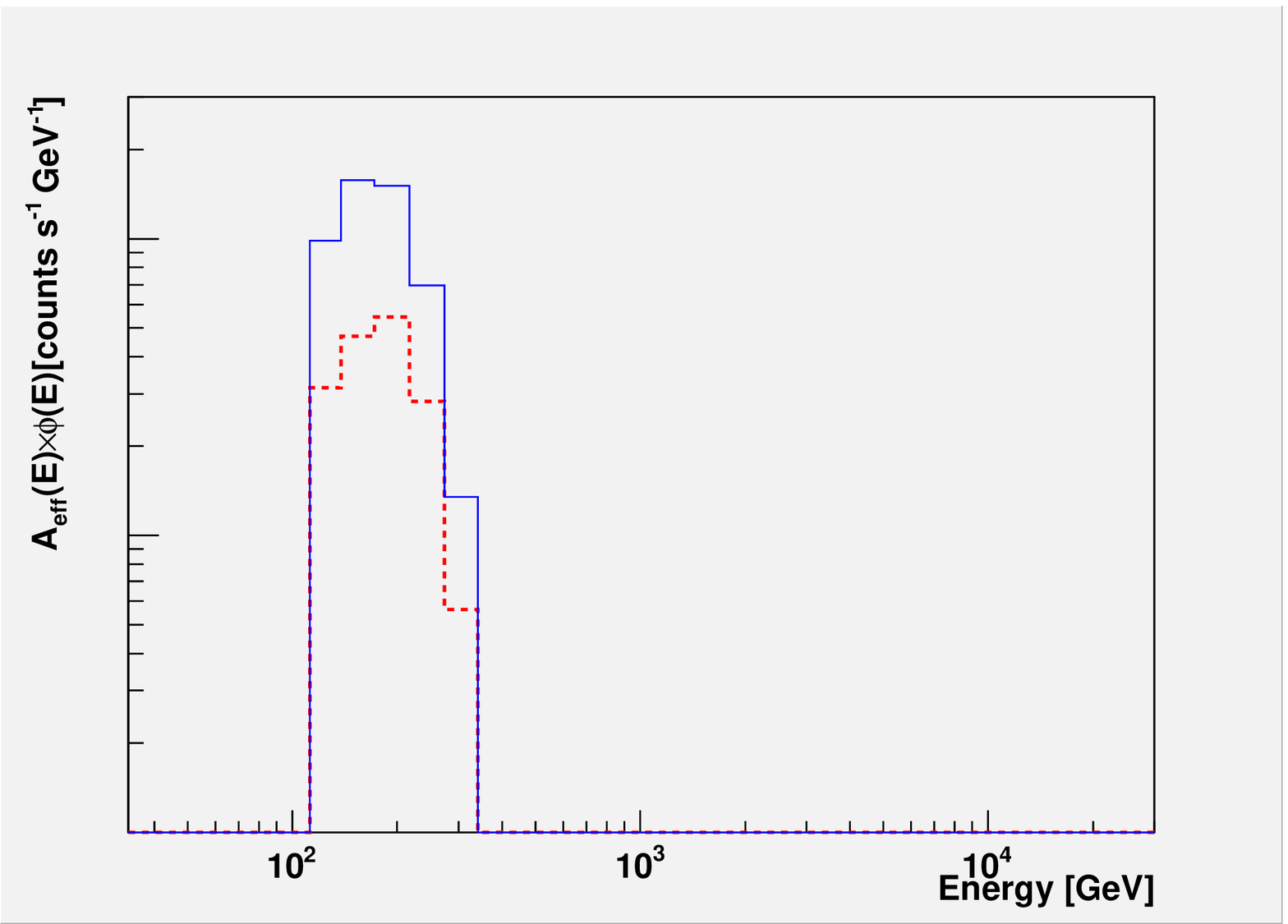}

	\caption{Response Functions for STACEE observations of Draco
		given two candidate spectra.  The left figure
		corresponds to an $E^{-2.2}$ spectrum, and the
		right figure corresponds to a Tyler spectrum
		(Eq.~\ref{tspec}) for an example
		$300~GeV/c^2$ WIMP.  The blue (solid) curves represent the
		STACEE sensitivity without cuts, the red (dashed) curves
 		include the grid-ratio cut.
	\label{detresp}}

	\end{center}
\end{figure*}

{\bf Power Law Spectrum:}
Figure~\ref{effarea} shows effective area curves for STACEE
observations of Draco.  Gamma-ray showers were simulated using the
Monte Carlo chain of the CORSIKA air shower simulation together with
our own optical ray-tracing model for the heliostats, secondaries, and
PMTs, and a simulation of the electronics~\cite{Lindner, Fortin}.  The
effective area is given by the product of the probability that a
shower triggers our detector and the area over which the simulated
showers were generated.

STACEE has an energy-dependent response which means the sensitivity to a
given source depends on its energy spectrum.
Figure~\ref{detresp} shows the result of convolving the effective area
curves with candidate spectra.  As is customary, we define the 
energy thresholds of our measurements as the peak of these curves.

The flux limit is defined by 
\begin{equation}
	N_{UL} = ~T~\int^\infty_0{ A(E)\Phi(E) dE }
\end{equation}
where $N_{UL}$ is given by the 95\% upper limit of the excess
$N_{ON}-N_{OFF}$, $T$ is the livetime, and $A(E)$ is the effective
area.  The differential flux, $\Phi(E) = C \phi(E)$, contains a
normalization constant with units of $[cm^{-2}~s^{-1}~GeV^{-1}]$.

\begin{figure}[tb]
	\begin{center}
	\includegraphics[width=0.45\textwidth]{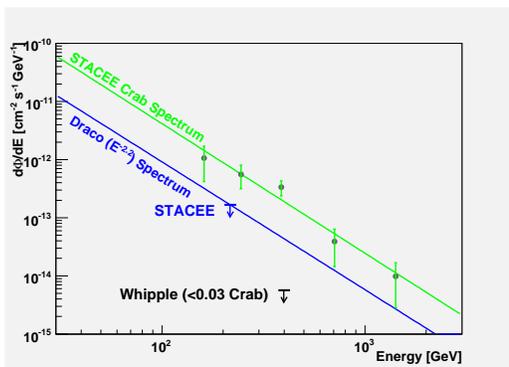}
	\caption{STACEE Flux limits for a $E^{-2.2}$ spectrum 
				as applied to Draco (blue).  For comparison, also shown is the energy spectrum of the Crab Nebula (green) as measured by STACEE during 2003-2005 which is well fit by the form: $\frac{dN}{dE}=1.2\times10^{-7}~E^{-2.23}$
	\label{flim}}
	\end{center}
\end{figure}

For the data given in Table \ref{datasum}, including
the grid-ratio cut, $N_{UL} = 138$ and the resulting upper limit is
\mbox{\bf $\Phi(E)<4\times10^{-8}~\left(\frac{E}{GeV}\right)^{-2.2}$}
\mbox{\bf $\gamma~s^{-1}~cm^{-2}~GeV^{-1}$}
at a characteristic energy of 220 GeV.
Figure \ref{flim} shows a comparison of this limit with the published
upper limit of the Whipple collaboration\cite{Hall} and our own Crab spectrum. 

\begin{figure}[tb]
	\begin{center}
	\includegraphics[width=0.45\textwidth]{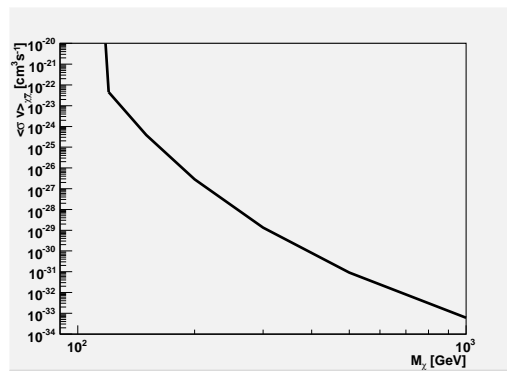}
	\caption{Upper limits on the WIMP self-annihilation rate
(cross-section multiplied by halo WIMP velocity) for the Tyler
spectrum as a function of $m_\chi$ as applied to the STACEE Draco
observations. 
	\label{dmlim}}
	\end{center}
\end{figure}

{\bf Tyler Spectrum:}
Tyler provides a prescription for converting a flux limit into a
cross-section limit (\mbox{$<\sigma v>_{\chi\bar\chi}$}) assuming a spherical
isothermal halo model where the mass density is given by $\rho_{halo} = Ar^{-2}$.  We avoid a divergence at the
center by including a constant-density core physically motivated by an
equilibrium between infalling particles and annihilation: 
\begin{equation}
	R_{min} = R_{ext} <\sigma v>^{1/2}_{\chi\bar\chi} 
		\left[ \frac{\rho_{halo}}{4 \pi G m_\chi^2} \right]^{1/4} 
\end{equation} 
where $R_{ext}$ is the outer radius of Draco
and $<\sigma v>_{\chi\bar\chi}$ is the expectation value of the
self-annihilation rate, given by the product of the cross-section and
the velocity of the WIMPs in the halo.
We then substitute this into our flux equation
\begin{equation}
	\Phi_\gamma(E) = \frac{4A^2}{3d^2R_{min}} <\sigma v>_{\chi\bar\chi}
	\phi_\gamma(E)
\end{equation}
and solve for the self-annihilation rate.  The resulting limits are shown in Figure
\ref{dmlim}.

\section{Conclusions}

STACEE does not detect a significant signal from Draco, and sets upper
limits on cross-sections for WIMP with rest-mass energy greater
than about 150 GeV.

{\bf Acknowledgments:}
Many thanks go to the staff of the National Solar Tower Test Facility, who
have made this work possible.
This work was funded in part by the U.S. National Science Foundation, the
Natural Sciences and Engineering Research Council of Canada, Fonds Quebecois
de la Recherche sur la Nature et les Technologies, the Research
Corporation, and the University of California at Los Angeles.

\bibliographystyle{unsrt}
\bibliography{icrc1076.bib}
\end{document}